\documentclass[doublecol]{epl2}
\usepackage{amsmath}
\usepackage{amsfonts}
\usepackage{dcolumn}
\usepackage{graphicx}
\usepackage{dcolumn}
\usepackage{color}
\usepackage{bm}
\usepackage{wrapfig}
\usepackage{sidecap}

\title{Spectral analysis of Gene co-expression network of Zebrafish}

\author{S. Jalan \inst{1} \and C. Y. Ung\inst{2} \and J. Bhojwani\inst{3}
\and B. Li\inst{4,5} 
\and L. Zhang\inst{6} \and S. H. Lan\inst{2} \and Z. Gong\inst{2}}

\institute{
\inst{1}School of Science, Indian Institute of Technology Indore, IET-DAVV Campus
Khandwa Road, Indore 452017, India\\
\inst{2}Department of Biological sciences, 
National University of Singapore 117546, Republic of Singapore\\
\inst{3}School of Life Sciences/Comp Sciences
DAVV, Indore\\
\inst{4}NUS Graduate School for Integrative Sciences and Engineering
117456, Republic of Singapore\\
\inst{5}Department of Physics and Centre for
Computational Science and Engineering, National University of Singapore
117546, Republic of Singapore
\inst{6}Department of Mathematics, National University of 
Singapore 117456, Republic of Singapore\\
}

\pacs{87.16.Yc}{Regulatory genetic and chemical networks}
\pacs{89.90.+n}{Other topics in areas of applied and interdisciplinary physics}

\abstract{
We analyze the gene expression data of Zebrafish under the combined framework of 
complex networks and random matrix theory. The nearest neighbor spacing 
distribution of the corresponding matrix spectra follows random matrix 
predictions of Gaussian orthogonal statistics. Based on the eigenvector analysis 
we can divide the spectra into two parts, first part for which the eigenvector 
localization properties match with the random matrix theory predictions, and the 
second part for which they show deviation from the theory and hence are useful 
to understand the system dependent properties. Spectra with the localized 
eigenvectors can be characterized into three groups based on the eigenvalues. We 
explore the position of localized nodes from these different categories. Using 
an overlap measure, we find that the top contributing nodes in the different 
groups carry distinguished structural features. Furthermore, the top 
contributing nodes of the different localized eigenvectors corresponding to the 
lower eigenvalue regime form different densely connected structure well 
separated from each other. Preliminary biological interpretation of the genes, 
associated with the top contributing nodes in the localized eigenvectors, 
suggests that the genes corresponding to same vector share common features.
}

\begin{document}

\maketitle

\section{Introduction}
Gene expression information captured in microarrays data for a variety of environmental and genetic 
perturbations 
promises to yield unprecedented insights into the organization and functioning of biological systems 
\cite{Lopez-Maury,Gasch,p-p1,lit}. 
The challenge no 
longer lies in obtaining gene expression profile, but rather in interpreting the results to gain insight into 
biological mechanisms. 
It has been increasingly realized that dissecting the genetic and chemical 
circuitry prevents us from further understanding the biological processes as a whole. In order to understand the 
complexities involved, all reactions and processes should be analyzed together. To this end, network theory has 
been getting fast recognition to study systems which could be defined in terms of units and interactions among 
them \cite{BA,rev-Strogatz,rev-network}. In this 
view one approach is to study the co-expression of genes, and to build up gene-sets working together. 
A gene 
co-expression network is defined by a set of nodes corresponding to genes, and a list of edges corresponding to 
co-expression. 
Using gene co-expression to recover co-regulated genetic modules is 
a standard approach adapted in system biology \cite{Stuart}.
We utilize gene expression data from Zebrafish exposed to various toxicants as study model \cite{data_gene2}.
The Zebrafish is an increasingly 
popular model not only for vertebrate development \cite{Grunwald} but also for understanding human diseases 
\cite{Shin} and toxicology \cite{Spitsbergen}. We analyze the gene co-expression network constructed from
Zebrafish data under the random
matrix theory (RMT) framework.

RMT was
initially proposed to explain the statistical properties of nuclear spectra \cite{mehta}. Later this theory was
successfully applied in the study of the spectra of different complex systems such as disordered systems, quantum
chaotic systems, and large complex atoms \cite{rev-rmt}. Further studies illustrate the usefulness of RMT in
understanding the statistical properties of the empirical cross-correlation matrices appearing in the study of
multivariate time series of price fluctuations in the stock market \cite{rmt-stock}, EEG data of
brain \cite{rmt-brain}, variation of various atmospheric parameters \cite{rmt-santh}, etc. Recent analysis of
complex networks under RMT framework \cite{pre2007a,pre2009a,pre2009b,rmt_aminoacid_prl2009} shows that various
model networks as well as real world networks follow universal GOE statistics. The analysis of protein-protein
interaction network of budding yeast reveals that
the nearest neighbor spacing distribution (NNSD) of the spectra of the corresponding matrix follows RMT
prediction \cite{pre2007a}. This promising result suggests that these networks can be modeled
as a random matrix chosen from an appropriate ensemble. 
Recently, covariance matrix of amino acid displacement \cite{rmt_aminoacid_prl2009} and gene co-expression network \cite{data_gene} constructed using gene expression
profiles from human brains  \cite{pre2010a} have been analyzed under RMT
framework. These analyses also show that the bulk of eigenvalues of corresponding networks follows 
universal GOE
statistics of RMT.
The universal GOE statistics of eigenvalues fluctuations
can be understood as some kind of randomness spreading over the real and model networks \cite{epl2009a}.

In this letter, we analyze the gene co-expression data under the random matrix theory framework. 
We find that
the bulk of the spectra follows random matrix predictions of the GOE statistics. Rest part of the spectra
deviates from the universality. We explore the properties of eigenvector from this part of the spectra in detail. Particularly,
we study the localization behavior of the spectra of the underlying matrix, and investigate the structural properties
of the top contributing nodes in the localized eigenvectors. 
We introduce an overlap measure to understand the structural properties of the nodes 
which are picked up based on the spectral properties of the underlying network.

\section{Method and Techniques}

\subsection{Construction of gene co-expression network}
One key problem in constructing gene co-expression network is to detect truly co-expressed gene pairs from 
genomic gene expression data \cite{Streib2008}. Pearson product method has been used traditionally to calculate 
correlation between pairs of genes using different conditions. Pearson correlation coefficient (PCC) between two 
variables is defined as the covariance of the two variables divided by the product of their standard 
deviations:\\
\begin{equation}
    \rho_{X,Y}={\mathrm{cov}(X,Y) \over \sigma_X \sigma_Y} ={E[(X-\mu_X)(Y-\mu_Y)] \over \sigma_X\sigma_Y},
\end{equation}
The absolute value of PCC is less than or equal to 1. Correlations equal to 1 or -1 correspond to maximum 
correlation and maximum anti-correlation respectively. The Pearson correlation coefficient is symmetric: 
$\rho(X,Y) = \rho(Y,X)$. Since in the microarrays studies the number of samples
is often limited, it is very crucial 
to find the most robust method to construct a co-expression network with the lowest effect of the sample size. To 
handle small sample size problem, PCC is calculated following a re-sampling bootstrap approach \cite{bootstrap}. 
The co-expression network is constructed, where two 
genes have weighted link if the correlation coefficient for that pair is greater than the value 
0.5, and the $95\%$ bootstrap confidence 
interval contains only positive numbers \cite{note1}. 
The corresponding network has entries $one$ or $zero$ depending upon whether the link between $i$ and $j$ 
is present or absent.
The original data set \cite{data_gene2} has $4021$ genes, after applying the threshold value $th=0.5$, we get 
the largest connected network with 
$N=4016$ nodes. The network is weighted with the distribution lying between $0.5 \le |{w_{ij}}| \le 1$.

\subsection{RMT Techniques}
In the following we briefly describe some of the RMT techniques used in our investigation. We denote the 
eigenvalues of a network by $\lambda_i,\,\,i=1,\dots,N$, where $N$ is size of the network and $\lambda_1 < 
\lambda_2 < \lambda_3 < \dots < \lambda_N$.  The density distribution $\rho(\lambda)$  
follows semi-circular distribution for GOE statistics. In order to calculate spacing distribution of eigenvalues, one has to remove the
spurious effects due to the variation of spectral density and to work at constant spectral
density on the average.
Thereby, it is customary in RMT to unfold the eigenvalues by a
transformation $\overline{\lambda}_i = \overline{N} (\lambda_i)$, where
$\overline{N} (\lambda) = \int_{\lambda_{\mbox{\tiny min}}}^\lambda\,
\rho(\lambda^\prime)\, d \lambda^\prime$ is the averaged integrated eigenvalue
density \cite{mehta}. Since analytical form for $\overline{N}$ is not known, we
numerically unfold the spectrum by polynomial curve fitting. 
The nearest neighbor spacing 
distribution ($P(s)$), where $s_i=\overline{\lambda_{i+1}}-\overline{\lambda_i}$,  
of eigenvalues follows $P(s)=\frac{\pi}{2}s\exp \left(-\frac{\pi s^2}{4}\right)$ for GOE statistics.
The distribution of the eigenvectors components provides system dependent information. Let $E_l^k$ is the $l$th
component of $k$th eigenvector $E^k$. The eigenvector components of a GOE random matrix are Gaussian distributed random
variables. For this case, the distribution of $r=|E_l^k|^2$, in the limit of large matrix dimension, is given by the
Porter-Thomas
distribution \cite{casati}. The inverse participation ratio (IPR) provides information about the localization properties of the eigenvectors \cite{Haake}. 
The IPR of a eigenvector is defined as
\begin{equation}
I^k = \frac{ (\sum_{l=1}^{N} [E_l^k]^2)^2}{ \sum_{l=1}^{N} [E_l^k]^4}
\label{eq-IPR}
\end{equation}
The meaning of $I^k$ is illustrated by the
two limiting cases : (i) a vector with identical components $E_l^k \equiv 1/\sqrt{N}$ has $I^k =N$, whereas (ii) 
a vector, with one component $E_1^k=1$ and the remainders zero, has $I^k=1$. Thus, the IPR quantifies the 
reciprocal of the number of eigenvector components that contribute significantly. A vector with components 
following the Porter-Thomas distribution has $I^k \sim N/3$.

\section{Results}
\label{results}
\begin{figure}
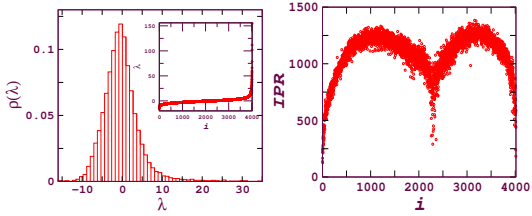

\vspace{-5pt}
\begin{center}
\includegraphics[width=0.39\columnwidth,height=2.7cm]{Fig1a.eps}
\includegraphics[width=0.39\columnwidth,height=2.8cm]{Fig1b.eps}
\end{center}
\vspace{-17pt}
\caption{(Color online) Left subfigure plots eigenvalue distribution for the largest connected cluster for 
the threshold value 
$th=0.5$. Inset plots eigenvalues in the increasing order. Right subfigure plots IPR as a function of eigen number $i$.}
\label{fig_eigen}
\vspace{-10pt}
\end{figure}

After applying the threshold we get the largest connected network with size $N=4016$ and $E=483148$ weighted edges. The weights are distributed between $-0.5$
and $0.5$. The 
average degree of this largest connected component is calculated as $<k> \sim 120$, and the degree distribution is 
shown to have exponential decay. The above number of edges yield a densely connected network. Higher values of 
threshold generate sparser networks, with number of nodes in the largest connected cluster lesser. The value of 
threshold $th=0.5$ is chosen in such a manner that it is sufficiently low to get almost all nodes into the largest 
connected cluster, and is sufficiently high to minimize the noise and measurement effect on the network 
\cite{threshold1,threshold2}. The bulk of the eigenvalues of this network lies roughly between $ 
-16$ and $30$. Last few eigenstates have steep increase in their values, with largest eigenvalue 
$\lambda_{max}=\lambda_N \sim 155$ being well separated from the bulk. The figure~\ref{fig_eigen} plots 
eigenvalues distribution, which shows triangular shape with exponential decay at both the ends. 
Using the unfolded spectrum, we calculate the
nearest neighbor spacing distribution $P(s)$, and find that
it follows GOE statistics of RMT. 

Figure~(\ref{fig_eigen}) shows IPR as a double humped well. For a network of size $N=4016$, the IPR value
for the eigenvectors following RMT prediction would be $\sim 1338$. Based on the eigenvector localization values, 
the spectra can be divided into two parts, one part with the delocalized eigenvectors having value 
close to the RMT prediction, 
and another part which consists the localized eigenvectors. According to RMT, this indicates that the corresponding 
network has a mixture of random connections yielding the delocalized eigenvectors of the first part, and the
structural 
features corresponding to functional performance leading to the localized second part. In order to get insight to the 
system dependent properties, we probe localized eigenvectors further. Based on the corresponding eigenvalues,   
the localized part of the spectra can further be divided into three distinct sub-parts, which we would discuss 
in detail. The first localized part (A) is associated with the lower eigenvalues regime, the second localized
part (B) corresponds to the middle
part of the spectra near the zero eigenvalue, and the third localized part (C) 
corresponds to the eigenstates with larger eigenvalues.

We make following general observations: the eigenvectors belonging to the part (A), in general, have 
the top contributing nodes with high degrees. The eigenvectors belonging to the part (B) have as few as one or 
two top contributing nodes. Additionally, these top contributing nodes have as few as one or two degrees. The 
eigenvectors belonging to the part (C) have top contributing nodes with degree close to the average degree of the 
network. The eigenvectors belonging to the part (C) do not have distinguished nodes contributing much higher than the
rest. Note that for finite dimensional matrix, deviation from randomness determines
the localization length of the eigenvectors \cite{Rev_Mod_Phy}.
\begin{figure}
\vspace{-4pt}
\centerline{\includegraphics[width=0.9\columnwidth]{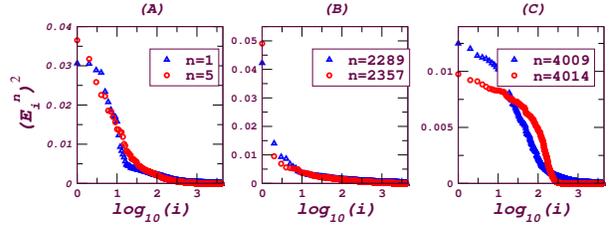}}
\vspace{-8pt}
\caption{(Color online) Typical behavior of the eigenvector elements from sets (A), (B) and (C). $n$ denotes the 
eigenstate, and $|E_i^n|^2$ denotes the $ith$ component of the $nth$ eigenvector. Nodes are reordered such that 
first top 
contributing node, where contribution is measured by $(E_i^n)^2$, gets index $i$ being 1, and last contributing 
node gets index $i=N$.}
\label{fig_eigenvec}
\vspace{-10pt}
\end{figure}
Figure~(\ref{fig_eigenvec}) plots the square of the eigenvector components for few eigenvectors lying at the bottom of 
the $IPR$ values. The different nature of eigenvector components in these three parts can clearly be seen. 
The eigenvectors from set (A) show the localization on approximately ten to twelve nodes, which contribute 
to the IPR highly. In the reordered nodes, as shown in the Figure~(\ref{fig_eigenvec}),  contributions from the top 
ten-twelve nodes decay with the node number, and finally reach to a plateau with very small values of 
$(E_i^n)^2$ for rest of the nodes. The eigenvectors in set (B) are highly localized on few nodes, namely one or two 
nodes, rest of the nodes lie towards the bottom of the participation measured by $(E_i^n)^2$. For 
the eigenvectors in set (C), the top contributing nodes give approximately same amount of contribution to the 
participation measure. The top 50 localized 
eigenvectors, except few, correspond to the eigenstates with the set (A), i.e. with the negative eigenvalues. 
Note that the eigenstate, corresponding to the largest eigenvalue ($\lambda_{N}$), has exponential decay 
components, and is not localized to few nodes.

In the following we analyze the top contributing nodes of eigenvectors lying 
towards the bottom of the IPR table. 
First, we analyze these nodes under the network theory framework, i.e. based on their degree and position in the 
network. We use the overlap measure to probe further the structural properties of the nodes.
Following this, we would note some of their functional relevance.

The top contributing nodes corresponding to the eigenvectors in the set (B), listed in the 
Ref.~\cite{supp}, are few, and interestingly one node appears in most of the localized eigenvectors in this set. The first localized 
eigenvector corresponds to the eigenstate $n=2289$. It has one localized node well separated from the others. 
This localized node comes at the number $3884$ in the largest connected component. The node lies to the periphery of the network with only one neighbor 
connected to it. The neighboring node $\mathcal{N}_{3884} \in \{1185\}$ has a very large degree $k_{1185} = 145$. 
The symbol $k_i$ denotes the degree of the $i$th node, and $\mathcal{N}_i \in \{j_1,j_2....j_k\}$ tells that the 
node $i$ has $k_i$ neighbors, namely, node $j_1$, node $j_2$ and so on.

The next localized eigenvector of set (B) corresponds to the eigenstate $n=2357$, and is localized on two nodes 
$n=4011$ and $n=4016$. The node number $4011$ is connected to only one node $3038$ with the degree $k_{3038} = 19$. 
The neighbors of $n=3038$ has varying degrees ranging from as low as $1$ to as high as $258$. The overlapping between 
all the pairs of nodes from the set of $\mathcal{N}_{3038}$ are less or equal to the overlapping in the corresponding 
random network. Node number $n=4016$ is connected with two other nodes $\mathcal{N}_{4016} \in \{3806, 3971\}$ 
with degrees $30$ and $13$ respectively. Furthermore, these neighbors also do not have any common neighbors.

In order to quantify the overlapping between neighbors of two nodes $i$ and $j$, here we define
a measure as,
\begin{equation}
O_{ij} = \frac{2 NN_{ij}}{min(NN_i,NN_j)}
\label{NN_Ratio}
\end{equation}
Where $NN_i$ is the number of neighbors of node $i$, and $NN_{ij}$ indicates the number of common neighbors 
between the nodes $i$ and $j$. Equation~\ref{NN_Ratio} measures the fraction of the overlapping neighbors of nodes 
$i$ and $j$.  The value $O_{ij} = 0$ corresponds to the case when the nodes $i$ and $j$ do not share any 
common neighbor, and the value $O_{ij}=1$ corresponds to the situation when all the neighbors of the node
$i$ ($j$) are within the set of neighbors of the node $j$ ($i$).

First, we calculate the overlap $O_{ij}$ between all the pairs of nodes $i$ and $j$, where $i$ is the neighboring 
node $3038$ of the first top contributing node $n= 4011$, and the node $j$ being any of the two neighbors $3806, 3971$ of 
the second top contributing node $4016$. The zero value of $O_{ij}$ tells there is no overlapping even between
the next to the nearest neighbors of the two top contributing nodes from the first localized eigenvector of set B. 

The next two localized eigenvectors of the set B, corresponding to eigenstates $n=2258$ and $n=2282$, also have one 
localized node $3884$ well separated from the others. The next two eigenvectors are localized on two nodes. 
Additional to the node number $3884$ which is the top contributing in the previous eigenvectors, there are
two other nodes, $3008$ and $4011$, appearing 
for the eigenvectors corresponding to $n=2298$ and $n=2307$ states respectively.

The localized eigenvectors from set (B) show entirely different features than the eigenvectors belonging to other 
two sets (A) and (C). The eigenvectors in this set not only have as few as one or two localized nodes, but also these nodes 
have few number of distinct neighbors.

\begin{figure}
\centering
\includegraphics[width=0.45\columnwidth]{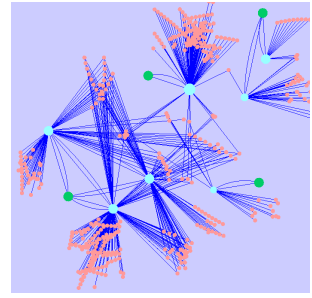}
\caption{(Color online) Part of the network consisting nodes and neighbors of the nodes corresponding to the set (B), see text for 
details. The top contributed nodes of the localized eigenvectors of this set are denoted by dark gray solid 
circles (green). The first neighbors of these nodes are denoted by light gray solid circles (cyan), and the second
neighbors are denoted by the light gray dots (orange).} 
\label{fig_Net_SetB} 
\vspace{-10pt}
\end{figure}

Figure~(\ref{fig_Net_SetB}) shows the section of the network consisting of the top contributing nodes,
of the localized eigenvectors of set (B). The sub-graph shows all the first (light gray solid circles, cyan) as well as 
second (light gray dots, orange, to the end of all links) neighbors of these nodes. 
The node number $3008$ (dark gray, green, solid circle towards bottom 
left) has three neighbors $\mathcal{N}_{3008} \in \{293, 302, 1821\}$ (light gray, cyan, solid circles connected 
with the node number $3008$) with large degrees.  The first pair of $\mathcal{N}_{3008}$ has 30 \% overlapping 
neighbors whereas other two pairs have $\sim 10\%$ and $\sim 20\%$ overlapping neighbors. Further more, we 
calculate $O_{ij}$ between all the pairs of $i$ and $j$, where $i$ denotes the neighbors of the first 
contributing node $n=3884$, and $j$ denotes the neighbors of the second contributing node $n=3008$. 
The values of $O_{ij}$ for such pairs are close to the value of $O_{ij}$ for the corresponding random network. 
Similarly, the two top contributing nodes from the eigenvector $E^{(2307)}$ have no next to the nearest 
common neighbors.

All the above observations suggest the followings: the top contributing nodes in the eigenvector from set (B) 
are either located on the periphery of the network, or they serve as a bridge to the several loosely separated 
communities. The top contributing nodes from the same eigenvectors from set (B) belong to the well separated 
different communities.

The top contributing nodes from set (A), which consists most of the localized eigenvectors, form a separate set 
for each eigenvector. The top most localized eigenvector corresponds to the minimum eigenvalue ($n=1$). Ten top 
most contributing nodes for this eigenvector, in serial, are: $ 99, 1101, 64, 129, 245,1243,1238,17, 91,226$. All 
of these nodes have very high degrees, and each node separately form clique of order 10 with its neighbors.

\begin{figure}
\centerline{\includegraphics[width=0.7\columnwidth]{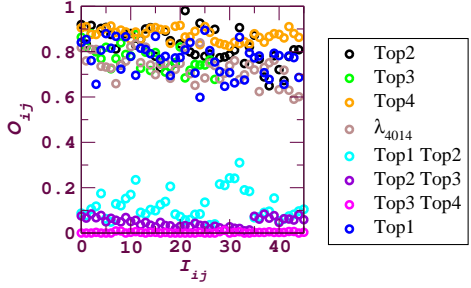}}
\vspace{-10pt}
\caption{(Color online) Fraction of overlapping neighbors $O_{ij}$ for the pairs of nodes $i$ and $j$. The horizontal
label shows the pair index $I_{ij}$, $i$ and $j$ denoting two different nodes, and the vertical axis plots the
overlapping between the nodes 
$i$ and $j$. The overlapping measure for the nodes within the eigenvectors $E^1, E^2, E^3, E^4$ and $\lambda_{max}$ 
are plotted with the different gray shades towards the top of the figure, and the overlapping measure for the 
pairs of nodes belonging to the different eigenvectors are plotted with different gray shades towards the bottom of 
the figure.
}
\label{fig_NN_Ratio}
\end{figure}

In order to measure the overlapping between the neighbors of the top contributed nodes, we calculate $O_{ij}$ for 
each pair of the nodes appearing in the top ten list of the different eigenvectors. Figure(~\ref{fig_NN_Ratio}) plots 
the overlapping measure $O_{ij}$ for the four most localized eigenvectors. As can be seen from the figure that 
the values of $O_{ij}$ for the pair of neighbors of the nodes from $E^{(4016)}$ are very high. The
overlapping for the pair of nodes beyond ten are inconsistent.  
Note that the network has the size $N=4016$ and the average degree $<k> \sim 120$, the number of common neighbors for a 
pair of nodes $i$ and $j$ in a corresponding random network would be of the order of $O_{ij} \sim 0.03$. 
Figure~(\ref{fig_NN_Ratio}) shows that the overlapping between neighbors of the top ten contributing nodes of 
eigenvector $E^{(4016)}$ is an order of magnitude larger than that of the corresponding random network. The
second most localized eigenvector corresponds to the eigenvalue $\lambda_5$. The top ten most contributing nodes 
again have degree towards the higher side, and they form completely different set than the top contributing nodes 
in the previous eigenvector. These nodes also form a clique of the order ten. Figure~(\ref{fig_NN_Ratio}) shows 
that the overlapping $O_{ij}$ between all the pairs of the top ten contributing nodes are very high. Similarly, 
the third and the fifth most localized eigenvectors also have completely different sets of the top ten 
contributing nodes. Note that the fourth most localized eigenvector 
belongs to the set (C). Figure~(\ref{fig_NN_Ratio}) plots $O_{ij}$ for all the pairs of top ten nodes corresponding to 
the fourth eigenvector as well.

Furthermore, $O_{ij}$ between the nodes from the different top eigenvectors for the set (A) are very low, 
Figure~(\ref{fig_NN_Ratio}) shows that the overlapping between the different pairs of the nodes from the
top four eigenvectors 
are of the order of the corresponding random network. Note that the degree of these nodes are very high, but they 
form different subgroups corresponding to each eigenvector, without significant common nodes even with the neighbors 
of the nodes from the different eigenvectors.

Furthermore, overlapping between the neighbors of the nodes belonging to the different localized eigenvectors 
from set (A) are few. The values of $O_{ij}$, where $i$ is a node of one eigenvector say $E^{(5)}$ and $j$ is 
a node from another eigenvector of set A, say $E^1$, comes in the bottom of the figure~\ref{fig_NN_Ratio}. All this 
suggests that the top contributing nodes in the different eigenvector from set (A) form densely connected community 
structure within a eigenvector, and which is loosely connected with the nodes from the other eigenvectors.

The first most localized eigenvector in the set (C) corresponds to the $\lambda_{4014}$. Though for this set, not 
much distinct localized nodes exist, the top ten contributing nodes have their degrees near to the average 
degree of the network, and these nodes are densely connected with each other. The very high values of $O_{ij}$ 
in the Figure~(\ref{fig_NN_Ratio}) for $\lambda_{4014}$ indicate that, all the pairs of top contributing nodes 
share large number of 
neighbors. These nodes form entirely different set than the top ten nodes of the eigenvector corresponding to the 
maximum eigenvalue $\lambda_{max}$. The eigenvector corresponding to the maximum eigenvalue comes at the number 
of $44$ in the terms of localization. Most of the top ten contributing nodes for this eigenvector are same as those for 
the first localized eigenvector which belongs to the set (A). Contrary to the top nodes in the other eigenvectors 
of the set (C), the top contributing nodes for $E^{(N)}$ have degree much larger than the average degree of the 
network. These nodes form a dense community with as high as $\sim 80\%$ of overlapping neighbors between each 
pair of the nodes. The second most localized eigenvector from this set (C) corresponds to the eigenstate $n=4009$. 
The average degree of the top ten contributing nodes is close to the average degree of the network, and the nodes
show very high overlapping between the neighbors.

\section{Conclusion and Discussion}
\label{conclusion}
We analyzed spectra of the gene co-expression network of Zebrafish generated for
different environmental perturbations. 
The eigenvalues statistics of the corresponding matrix shows triangular distribution with exponential decay at both 
the ends. Triangular distribution is one of the known characteristics for scalefree networks. The 
spacing distribution of the eigenvalues follows GOE statistics of RMT, which indicates that there is 
{\it sufficient amount of randomness} \cite{epl2009a} existing in the co-expression network. The 
eigenvector localization measured by 
the IPR values shows that the bulk of the spectra follows RMT predictions of the GOE statistics. The remaining 
part of the spectra which deviates from the RMT predictions carries system dependent information. 
The later 
part of the spectra have localized eigenvectors, and is divided into three groups
based on the associated eigenvalues. Moreover, top contributing nodes corresponding to the different
groups show different structural features.
The eigenvectors 
belonging to the first group (A) has top contributing nodes with high degrees, and are associated with the 
lower eigenvalues regime. The eigenvectors associated with the second group (B) have top contributing nodes with 
as few as one or two degrees, and are associated with the zero eigenvalue regime. The eigenvectors from group (C) 
lie towards the largest eigenvalue, and have top contributing nodes with degrees close to the average degree of the 
network. Most of the 
top localized eigenvectors belong to the part (A), i.e. correspond to the lower eigenvalues regime. According to 
the RMT, the localized eigenvectors distinguish 'genuine correlations from apparent correlations' 
\cite{rmt-stock} which, in terms of the gene co-expression networks, 
can be interpreted as {\it random correlations} between the genes and {\it 
functionally important correlations}
between the genes. The corresponding matrix is not random, and hence leads to the localization of some of the
the eigenvectors. The top contributing nodes in the localized eigenvalues may have important structural and 
functional roles.

In order to probe the structural relevance of the top contributing nodes we define overlap measure. For
the set (B), overlapping measure shows that the top contributing nodes lie 
towards the periphery of the network. 
Additionally, based on the overlap measure 
for the neighbors of the top contributing nodes from localized eigenvectors, we see that these nodes belong 
to the different regions of the network with a small overlapping between
even the next to the nearest neighbors. The 
largest overlap in these pairs is those of the corresponding random networks. 
These observations suggest that for the set (B), 
the top contributing nodes belonging to same eigenvectors lie either near
to the periphery, or belong to the 
different parts of the networks.

For the set (A), the top contributing nodes of a eigenvector have very high values of the overlap measure, infact 
for some of the localized eigenvectors these nodes are part of the clique of the order ten. Furthermore, the top 
contributing nodes from the different localized eigenvectors lie well separated from each other. All the 
observations for set (A) 
indicate that the top contributing nodes in different eigenvectors form separate community structure; these 
communities are densely connected within, and are loosely connected with each other.

The top contributing nodes from the set (C) do not show any noticeable structural features, except the fact that, the nodes 
from each localized eigenvector form densely connected set though they have less number of neighbors, equal to 
the average degree of the network.

Description in the Ref.~\cite{supp} suggests that most of the genes in the localized sets, which are functionally related or are 
members of pathways leading to similar diseases, are basically clustered in one set. The exceptional cases, 
however, may indicate that although these genes were identified in the same set, having different function 
altogether, could have similar protein expression patterns that are regulated by similar transcriptional cues in 
similar developmental domains, like that of the other members in that particular set. Though the biological 
interpretation drawn in the letter is at a preliminary stage, the analysis presented here shows the 
applicability and the usefulness of random matrix theory to pick out set of nodes (genes) from a large number of 
interacting nodes (genes), which were not detected based on existing structural measures. Identification of 
genes that are significantly responsible for diverse
toxicological perturbations is important in order to develop a future chip, that can be used for detection of pollutants        
and diagnosis of diseases. 
Future directions would involve hand in hand experiments and theoretical 
investigations to make direct relation between cause such as environmental perturbations and gene or set of genes 
affected.

\section{Acknowledgment}
The microarrays works were supported by Environment and Water Industry Grant of Singapore. CYU and LZ are
currently supported by grant MOE2009-T2-2-064. SJ acknowledges DST grant SR/FTP/PS-067/2011
for the financial support.


\begin{thebibliography}{99}
\bibitem{note1}If instead of the bootstrapping approach we use just the sample 
correlation coefficient, a larger set of links would be present which would contain genes having very large 
bootstrap confidence intervals.
\end{thebibliography}
\end{document}